\documentclass[a4paper,11pt]{article}

\usepackage[dvips]{graphicx}
\usepackage{amsmath,amssymb}
\usepackage{color}

\usepackage[normalem]{ulem}

\makeatletter
\renewcommand\@biblabel[1]{}
\makeatother

\begin{document}

\author{Adam B.~Barrett\footnote{adam.barrett@sussex.ac.uk}\\\\
\textit{Sackler Centre for Consciousness Science} and \textit{Department of Informatics}\\
University of Sussex, Brighton BN1 9QJ, UK}

\title{An Integration of Integrated Information Theory with Fundamental Physics}

\date{[Published Feb.~4, 2014 in the \textit{Consciousness Research} specialty section of \textit{Frontiers in Psychology}, article no.~5(63).]}

\maketitle

\begin{abstract}
\noindent To truly eliminate Cartesian ghosts from the science of consciousness, we must describe consciousness as an aspect of the physical. Integrated Information Theory states that consciousness arises from intrinsic information generated by dynamical systems; however existing formulations of this theory are not applicable to standard models of fundamental physical entities. Modern physics has shown that fields are fundamental entities, and in particular that the electromagnetic field is fundamental. Here I hypothesize that consciousness arises from information intrinsic to fundamental fields. This hypothesis unites fundamental physics with what we know empirically about the neuroscience underlying consciousness, and it bypasses the need to consider quantum effects.
\end{abstract}

\section*{Introduction}

The key question in consciousness science is: ``Given that consciousness (i.e., subjective experience) exists, what are the physical and biological mechanisms underlying the generation of consciousness?''. From a basic property of our phenomenology, namely that conscious experiences are integrated representations of large amounts of information, Integrated Information Theory (IIT) hypothesizes that, at the most fundamental level of description, consciousness is integrated information, defined as information generated by a whole system, over and above its parts (Tononi, 2008). Further, given the private, non-externally observable nature of consciousness, IIT considers consciousness to be an intrinsic property of matter, as fundamental as mass, charge or energy. Thus, more precisely, IIT posits that consciousness is intrinsic integrated information, where by intrinsic information it is meant that which is independent of the frame of reference imposed by outside observers of the system. The quantity of consciousness generated by a system is the amount of intrinsic integrated information generated (Balduzzi and Tononi, 2008), whilst the qualities of that consciousness arise from the precise nature of informational relationships between the parts of the system (Balduzzi and Tononi, 2009).

IIT has garnered substantial attention amongst consciousness researchers. However, it has been criticized for its proposed measures of integrated information not successfully being based on an intrinsic perspective (Gamez, 2011; Beaton and Aleksander, 2012; Searle, 2013). The proposed ``$\Phi$'' measures are applicable only to networks of discrete nodes, and thus for a complex system depend on the observer choosing a particular graining. More broadly, information can only be intrinsic to fundamental physical entities, and descriptions of information in systems modeled at a non-fundamental level necessarily rely on an extrinsic observer's choice of level (Floridi, 2009, 2010; Gamez, 2011). Here I propose a potential solution to this problem, what might be called the field integrated information hypothesis (FIIH). Modern theoretical physics describes the universe as being fundamentally composed of continuous fields. Electrical signals are the predominant substrate of information processing in brains, and the electromagnetic field that these produce is considered fundamental in physics, i.e., it is not a composite of other fields. Thus, I hypothesize that consciousness arises from information intrinsic to fundamental fields, and propose that, to move IIT forward, what is needed is a measure of intrinsic information applicable to the configuration of a continuous field.

The remainder of this article is laid out as follows. First I discuss the concept of fundamental fields in physics, and how if one takes the view that consciousness is an intrinsic property of matter, then it must be a property arising from configurations of fields. In the following section, I discuss the hypothesis that consciousness arises from integrated information intrinsic to fundamental fields, the shortcomings of existing approaches to integrated information, and the possibility of constructing a measure that can successfully measure this quantity for field configurations. I then explain how IIT and the FIIH imply a limited form of panpsychism, and why this should not be considered a problem, before contrasting the FIIH with previously proposed field theories of consciousness, such as that of Pockett (2000). Finally, the summary includes some justification for this theoretical approach to consciousness.

\section*{Fundamental fields and consciousness}

\begin{table}[h!]
\begin{center}
{
\begin{tabular}{lllll}
\hline
 & \textbf{Mass (GeV/$c^2$)} & \textbf{Electric charge} & \textbf{Strong charge} & \textbf{Weak charge} \\
\hline
\hline
\textbf{LEPTONIC MATTER}   & & & &  \\
\hline
electron neutrino ($\nu_e$) & $<1.3\times10^{-10}$ & 0 &  No & Yes\\
electron (e) & 0.0005 & -1 &  No & Yes \\
\hline
muon neutrino ($\nu_\mu$) & $<1.3\times10^{-10}$ & 0 &  No & Yes\\
muon ($\mu$) & 0.106 & -1 &  No & Yes\\
\hline
tau neutrino ($\nu_\tau$) & $<1.4\times10^{-10}$ & 0 &  No & Yes\\
tau ($\tau$) & 1.78 & -1 &  No & Yes\\
\hline
\hline
\textbf{QUARK MATTER}  & & & & \\
\hline
 up (u) & 0.002 & 2/3 &  Yes & Yes\\
down (d) & 0.005 & -1/3 & Yes & Yes\\
\hline
charm (c) & 1.3 & 2/3& Yes & Yes \\
strange (s) & 0.1 & -1/3 & Yes & Yes\\
\hline
top (t) & 173 & 2/3 & Yes & Yes\\
bottom (b) & 4.2 & -1/3 & Yes & Yes\\
\hline
\hline
\textbf{BOSONS}  & & & & \\
\hline
\textbf{Electromagnetic force:} & & & & \\
photon ($\gamma$) & 0 & 0 & No & No \\
\hline
\textbf{Strong force:} & & & & \\
gluon (g) & 0 & 0 & Yes & No \\
\hline
\textbf{Weak force:} & & & & \\
$W^-$ & 80 & -1 & No & No \\
$W^+$ & 80 & 1 & No & No \\
Z & 91 & 0  & No & No \\
\hline
\textbf{Gravity:} & & & & \\
graviton$^*$ & 0 & 0  & No & No \\
\hline
\textbf{Higgs mechanism:} & & & & \\
Higgs (H) & 126 & 0 & No & Yes \\
\hline
\\
\end{tabular}
}
\caption{Table of the fields/particles that are considered fundamental. Familiar matter arises from leptons and quarks, while the forces of nature arise from interactions of matter with ``carrier'' bosons. Mass is given in giga electron volts per speed of light squared (Gev/$c^2\approx 2\times10^{-27}$kg). Electric charge is in standard units relative to minus the charge of the electron, i.e., one unit equals $1.6\times10^{-19}$ Coulombs. A description of the group theoretic strong and weak charges is beyond the scope of this article, but the table shows which fields have strong and weak charges. *The gravity field is considered fundamental and is well-studied, but the gravity particle (graviton) has not to date explicitly been observed; at quantum (i.e., very microscopic) spatial scales, a consistent set of field equations for gravity have yet to be constructed.}
\end{center}
\vspace{0.5cm}
\end{table}

Contemporary physics postulates that ``fields'' are the fundamental physical ingredients of the universe, with the more familiar quantum particles arising as the result of microscopic fluctuations propagating across fields, see e.g., Oerter (2006) for a lay person's account, or Coughlan et al.~(2006) for an introduction for scientists. In theoretical terms, a field is an abstract mathematical entity, which assigns a mathematical object (e.g., scalar, vector) to every point in space and time. (Formally a field is a mapping $F$ from the set $S$ of points in spacetime to a scalar or vector field $X$, $F: S \to X$.) So, in the simplest case, the field has a number associated with it at all points in space. At a very microscopic scale, ripples, i.e., small perturbations, move through this field of numbers, and obey the laws of quantum mechanics. These ripples correspond to the particles that we are composed of, and there is precisely one fundamental field for each species of fundamental particle. At the more macroscopic level, gradients in field values across space give rise to forces acting on particles. The Earth's gravitational field, or the electromagnetic field around a statically charged object, are examples of this, and the classical (as opposed to quantum) description is a good approximation at this spatial scale. However, both levels of description can be considered equally fundamental if the field is fundamental, i.e., not some combination of other simpler fields. Note that the electromagnetic and gravitational fields are both examples of fundamental fields, with the corresponding fundamental particles being the photon and the graviton. Particles are divided up into matter particles and force-carrying particles, but all types of particle have associated fields; all the forces of nature can be described by field theories which model interactions, i.e., exchanges of energy, between fields. See Table 1 for a list of fields/particles that are considered fundamental according to this so-called ``Standard Model'' of particle physics.

To be consistent with modern theoretical physics, a theory of consciousness that considers consciousness to be a fundamental attribute of matter must describe how consciousness manifests itself in the behavior of either fundamental fields or quantum particles. Since we know that the brain generates electric fields with a rich spatiotemporal structure, and that, for the main part, information processing in the brain is carried out by electrical signaling between neurons operating mostly in the classical (as opposed to quantum) regime (Koch and Hepp, 2006), empirical evidence favors the former. Thus, on the view that consciousness is a fundamental attribute of matter, it must be the structure and/or dynamics of the electromagnetic field (which is an example of a fundamental field) that is fundamentally the generator of brain-based consciousness.

Once one ascribes electromagnetic fields with the potential to generate consciousness, it is natural to ask whether other fields might also have the potential to generate consciousness. According to modern physics, there was a symmetry between all fields at the origin of the universe, although these symmetries were broken as the universe began to cool (Georgi and Glashow, 1974; see Hawking, 2011 for a lay-person's account). It could be argued by Occam's razor that it makes more sense to posit that potential for consciousness existed at the outset, and hence potential for consciousness is a property of all fields, than that it emerged only during symmetry breaking. However, in practice, it is unlikely that any complex consciousness could exist in any field other than the electromagnetic field, for reasons to do with the physics and chemistry of the electromagnetic field compared with other fields. Considering the four forces: strong, weak, electromagnetic and gravitational, the strong and weak forces don't propagate over distances much larger than the width of the nucleus of an atom, and gravity alone cannot generate complex structures by virtue of being solely attractive; in contrast, the electromagnetic field can propagate over macroscopic scales, is both repulsive and attractive, and is fundamentally what enables non-trivial chemistry and biology. Considering fields associated with matter, these in general do not have any undulations at spatial scales larger than the quantum scale; the non-trivial structures in these fields are essentially just the ripples associated with the familiar quantum matter particles, i.e., electrons and quarks, and various ``exotic'' particles detectable in particle physics experiments (see Table 1). Finally, the recently discovered Higgs field has essentially a uniform structure; quantum interactions exist between the Higgs field and many of the other fields, and this is fundamentally the origin of mass in the universe (see e.g., Coughlan et al., 2006; Oerter, 2006). Thus, the physics of the electromagnetic field uniquely lends itself to the generation of complex structures.

\section*{The Field Integrated Information Hypothesis}

Given the above, I propose that the principal conceptual postulates of IIT should be restated as follows. Consciousness arises from information intrinsic to the configuration of a fundamental field. The amount of consciousness generated by a patch of field is the amount of integrated information intrinsic to it. When a patch of field generates a large quantity of intrinsic integrated information, mathematically there is a high-dimensional informational structure associated with it (Tononi, 2008; Balduzzi and Tononi, 2009). The geometrical and topological details of this structure determine the contents of consciousness. The task now is to correctly mathematically characterize intrinsic integrated information, and construct equations to measure it.

A true measure of intrinsic integrated information must be frame invariant, just like any fundamental quantity in physics. That is, it must be independent of the point of view of the observer: independent of the units used to quantify distance or time, independent of which direction is up, and independent of the position of the origin of the coordinate system; and also independent of the scale used for quantifying charge, or field strength.

The ``$\Phi$'' measures put forth by existing formulations of IIT (Balduzzi and Tononi, 2008; Barrett and Seth, 2011) are not applicable to fields because they require a system with discrete elements, and fields are continuous in space. One could ask, however, whether a perspective on a system in terms of discrete elements could actually be equivalent to an intrinsic field-based perspective, thus obviating the need for a field-based measure. To see explicitly that this is not the case, let us revisit the photodiode, which, according to the existing theory (Tononi, 2008), has 1 bit of intrinsic information by virtue of having two states, on or off. There is a wire inside the photodiode, and the electrons inside the wire are all individually fluctuating amongst many different states. The electromagnetic field generated by the diode, and the circuit to which it is connected has two stable configurations for as long as the circuit is connected. But other more general configurations for an electromagnetic field are ruled out by each of these states. Considering the system at this level of description yields a distinct perspective, and would lead one to deduce that the amount of information generated by the system's states is some quantity other than 1 bit. Thus the field-based perspective is not equivalent to the observer-dependent discrete perspective.

The idea here is that a formula should be obtained that could in theory be applied universally to explore the intrinsic information in any patch of spacetime, without requiring an observer to do any modeling, i.e., one would just measure field values in as fine a graining as possible to get the best possible approximations to the intrinsic informational structure. Only a formula in continuous space and time would allow this. If a discrete formula were to be applied, there would always be the possibility of encountering an informational structure on a finer scale than that of the formula. (Unless the graining required by the formula were the Planck scale, i.e., the scale of the hypothesized superstring, on which continuous models of physics break down; however there do not exist complex structures at that scale.) In practice however, observations of systems are necessarily discrete, so discrete approximations to a continuous formula could be useful for empirical application. See Balduzzi (2012) for some recent work on the information-theoretic structure of distributed measurements.

We don't yet know how to properly calculate intrinsic information, so must remain agnostic on the precise amount of intrinsic integrated information generated by photodiodes, or of anything. However, the failure of existing approaches does not rule out the construction in the future of a successful formula. While it is beyond the scope of this present paper to make a serious attempt at solving this problem, I speculate that a formula in terms of thermodynamic entropy as opposed to Shannon entropy might be more likely to succeed, as the former is inherently an intrinsic property, whereas the latter was constructed for the purpose of describing an external observer's knowledge of a system (Floridi, 2009, 2010; Gamez, 2011; Beaton and Aleksander, 2012).

\section*{Integrated Information Theory and panpsychism}
Searle (2013) criticizes IIT for its stance that integrated information always produces consciousness, stating that this ludicrously ascribes consciousness to all kinds of everyday objects and would mean that consciousness is ``spread thinly like a jam across the universe''. Koch and Tononi (2013) counter that only ``local maxima'' of integrated information exist (over spatial and temporal scales): ``my consciousness, your consciousness, but nothing in between''. If local maxima of intrinsic integrated information in field configurations always generate consciousness, then there must be minute amounts, say ``germs'', of consciousness all over the universe, even though there would be no superordinate consciousness amongst groups of people. Thus, IIT and the FIIH do imply a form of panpsychism. However, the phenomenology assigned to an isolated electron in a vacuum, or even a tree, which has no complex electromagnetic field, would be very minimal. Since the only consciousness we can be certain of is our own, the positing by integrated information theories of germs of consciousness everywhere is no reason to dismiss them. A theory should stand or fall on whether or not it can elegantly and empirically describe human consciousness.

For those uncomfortable with subscribing to a panpsychist theory, a possible way round the problem is to assign an attribute ``potential consciousness'' to matter at the most fundamental level. Then, the quantity of potential consciousness is simply the quantity of integrated intrinsic information. But only when there is a large amount of intrinsic integrated information with a sufficiently rich structure to be worthy of being compared to a typical healthy adult human waking conscious moment, should we say that the integrated information has ``actual consciousness'' associated with it. A line could thus be drawn somewhere between the potential consciousness of an isolated electron in a vacuum and the actual consciousness generated by my brain as I write this article. The problem with such a distinction however is that potential consciousness would still be assigned phenomenal content, so it is perhaps more elegant to just use a single term ``consciousness'' for the whole spectrum of integrated information. On the other hand, since consciousness is defined by some as any mental content, but by others as only self-reflective mental content, there is no single terminology that appeals to everybody. The key point, irrespective of the precise definition of consciousness, is that on the theory discussed here, intrinsic integrated information is what underlies subjective experience at the most fundamental level of description. Alternatively, one could further imagine different lines being drawn for different purposes. For example, a threshold of conscious awareness above which surgery cannot be performed; or thresholds at which various people are comfortable eating animals.

\section*{Relation to previous electromagnetic field theories of consciousness}
There have been several other theories of consciousness put forward that identify consciousness with various types or configurations of fields, see Pockett (2013) for a review. Notably, Pockett's electromagnetic field theory (EMT) of consciousness (Pockett, 2000, 2011, 2012) posits that ``conscious perceptions (and sensations, inasmuch as they can be said to have independent existence) are identical with certain spatiotemporal electromagnetic patterns generated by the normal functioning of waking mammalian brains'' (Pockett, 2013). In the most recent formulation of this theory, the key feature of field patterns underlying consciousness is the presence of a neutral region in the middle of a radial pattern. This hypothesis was motivated by the observation that such field patterns appear during recurrent cortical activity, (with the neutral region in layer 4), and the empirical association of consciousness with recurrent processing (Pockett, 2012).

A problem common to previous field theories of consciousness (Libet, 1994; Pockett, 2000, 2013; McFadden, 2002) is that they claim that cutting outgoing neural connections from a slab of cortex that generates a conscious experience will not affect the ability to report that conscious experience. EMT argues that the electromagnetic field within such an isolated hypothetical slab would still propagate through space and enable communication between the conscious field generated by the slab and the spatially contiguous larger conscious mental field. This is not however compatible with the laws of physics. Any cutting of synapses to or from regions of cortex that are generating consciousness will alter the field, and will therefore alter the conscious experience. There is no electromagnetic field residing in the brain other than that generated specifically by all of the neural and chemical activity. And it does not make sense to talk of the brain's electromagnetic field and its firing neurons and synapses as being able to exist independently of each other. On the theory put forward here, neurons can be considered the scaffolding that enable very complex electromagnetic field configurations to be sustained. As far as describing the mechanisms of perception and cognition that generate the specific contents of consciousness in any given scenario, the current paradigm of associating it with neural activity is of course the only valid and useful level of description. However, in terms of explaining more fundamentally how matter gives rise to consciousness, a description in terms of fields would be much more elegant than a description in terms of the complex entities that are neurons.

Another shortcoming of previous field theories of consciousness is that none of them relate physical properties of proposed correlates of consciousness to properties of phenomenology, i.e., they do not posit ``explanatory correlates of consciousness'' (Seth, 2009). The FIIH raises for the first time the possibility of constructing a field theory of consciousness that can account for a fundamental aspect of phenomenology, namely that conscious experiences are integrated representations of large amounts of information.

\section*{Discussion}

In this paper I have hypothesized that, at the most fundamental level of description, human consciousness arises from information intrinsic to the complex electromagnetic fields generated by the brain. This ``FIIH'' builds on the axioms of IIT, namely that consciousness is integrated information, and that consciousness is an intrinsic and fundamental property of matter analogous to mass or charge. However, it also implies that a new mathematical formalism is required to properly quantify intrinsic integrated information, since electromagnetic fields are continuous in space, and existing ``$\Phi$''-type measures of integrated information are applicable only to discrete systems (which require an observer dependent perspective). The idea that consciousness can be identified with certain spatiotemporal electromagnetic patterns has been previously put forward in other electromagnetic field theories of consciousness. But by suggesting that integrated information is the key factor, the theory here connects, for the first time, such electromagnetic field theories of consciousness to basic aspects of phenomenology.

The hypothesis is admittedly rather speculative, and any proposed mathematical formula for conscious level in terms of information intrinsic to an electromagnetic field will be difficult to test directly, simply because we do not have the technological tools or the computational resources to record in full detail the three-dimensional electromagnetic field structure generated by the brain. Rather, this can only be sampled at a spatial scale that is sparse compared to the finest scale of its undulations. However, there is a strong case to be made that the theoretical development of the ideas presented here has substantial value. Theories in physics have been vigorously pursued for their logic and beauty, in the absence of imminent direct experimental tests. For example, there is a vast amount of work being conducted on string theory; there, rather than experimental verification, the goal is an elegant explanation of our existing empirical knowledge of particle physics and gravity. If there already existed several analogous theories of consciousness, then one could argue that it would not be useful to add to the speculation. However, there is as yet no compellingly believable set of equations for describing, fundamentally, how consciousness is generated. IIT has potential in this direction, but a major step forward for the theory would be a truly plausible formula for intrinsic information applicable to fundamental physical entities. The FIIH provides a conceptual starting point for achieving this. All this is not to say that such a theory will aid understanding of all aspects of consciousness; indeed the multi-faceted nature of consciousness requires descriptions at many different levels. Non-reductionist frameworks are required to understand the complexity of the biological machinery that enables the brain to do any kind of information processing, conscious or unconscious, and to understand the differences between conscious and unconscious cognitive processes neural dynamics and behavior must necessarily be modeled at multiple levels of description.

Finally, any theory can potentially indirectly make predictions. Indeed IIT has already inspired heuristic measures of information integration/complexity that have been successfully applied to recorded electrophysiological data and are able to distinguish the waking state from diverse unconscious states, i.e., sleep and anaesthesia under various anaesthetics (Massimini et al., 2005; Ferrarelli et al., 2010; Casali et al., 2013). The results are in broad agreement with the predictions of IIT and provide encouragement for further theoretical work on the relationship between information integration and consciousness. Theories built from the FIIH could make new and distinct predictions about the types of structural and/or functional neuronal architectures that are capable of generating consciousness; and new theory can only further inform the quest for ever more reliable measures of consciousness that can be applied to observable brain variables.

\section*{Acknowledgements}
I thank Emily Lydgate and Anil Seth for invaluable discussions during the writing of this paper, and Daniel Bor and David Gamez for very useful comments on draft manuscripts. ABB is funded by EPSRC grant EP/L005131/1.

\bibliographystyle{frontiersinSCNS&ENG} 

\end{document}